\documentclass{aa}
\usepackage{graphicx}
\usepackage{txfonts}
\usepackage{natbib}
\bibpunct{(}{)}{;}{a}{}{,}

\newcommand{\gs}{\raisebox{-.8ex}{$\buildrel{\textstyle>}\over\sim$}}
\newcommand{\ls}{\raisebox{-.8ex}{$\buildrel{\textstyle<}\over\sim$}}

\begin{document}

\title
{On the growth and stability of Trojan planets}

\author{P.\ Cresswell and R.~P.\ Nelson}

\institute{ Astronomy Unit, 
 Queen Mary, University of London, Mile End Rd, London, E1 4NS, U.K.}
 
\offprints{P.Cresswell@qmul.ac.uk}

\date{Received /Accepted}

\def\LaTeX{L\kern-.36em\raise.3ex\hbox{a}\kern-.15em
         T\kern-.1667em\lower.7ex\hbox{E}\kern-.125emX}

\authorrunning{P.\ Cresswell \& R.~P.\ Nelson}
\titlerunning{Growth \& stability of Trojan planets}

\abstract
{}
{We investigate the stability of those low-mass Trojan planets that form in
a protoplanetary disc and subsequently accrete gas to become gas giants.}
{We calculate their evolution before, during, and after gas disc dispersal.
A two-dimensional hydrodynamics code combined with an $N$-body solver is
used to evolve the system of disc and planets. Gas disc dispersal
is simulated in a simple manner by assuming global exponential 
decay of the disc mass,  leading to the stalling of migration after
semi-major axes have approximately halved from their initial values. 
We consider Trojan pairs with different initial
masses and gas accretion rates and gas disc models with different masses
and viscosities.  An $N$-body code, adapted to model disc forces, is
used to examine large-scale migration and the formation of 
very short period Trojan planets.}
{For each combination of planetary pair and disc model that we consider
 in our hydrodynamic simulations, 
each Trojan system
remains stable before, during, and after disc dispersal. The long-term 
stability of these systems in the absence of gas 
is tested using $N$-body simulations, and all systems remain
stable for those evolution times equal to $10^9$ years. Eccentricities 
remain low ($e<0.02$) in all cases.
 Increases in the amplitude of libration about the $L_4$/$L_5$ Lagrange points 
accompany the inward migration, and during very large-scale migration
Trojan systems may be disrupted prior to the onset of disc dispersal.}
{The stability of Trojan pairs during rapid type I migration, during the
transition to type II migration with the accompanying gap formation in the
gas disc, and 
during gas loss when the disc disperses, 
indicates that isolated Trojan planet systems are very stable. 
If a common mechanism exists for their formation, we suggest they may be 
readily observed in nature. }


\maketitle

\section{Introduction}
\label{intro}

As we become aware of an ever-increasing number of exoplanetary systems
containing multiple planets, 
we find that the vast majority thus discovered all differ greatly in some 
manner from our own Solar System
\citep{fischer08, cochran07, udry07}.
Some potential planetary system architectures have been hypothesized but not 
yet observed; however, 
among these are the Trojan planets: planetary pairs orbiting a star 
with very similar semi-major axes but with a
leading/trailing separation in azimuth of $ \simeq 60^{\circ}$. 
Nonetheless, such a discovery would not be completely without precedent.
In our own solar system Jupiter and Neptune both possess sizeable
populations of Trojan asteroids at their $L_4$ \& $L_5$ points, while some satellites
of the giant planets lie in co-orbital horseshoe (Janus \& Epimetheus) and
tadpole (Dione, Helene and Polydeuces) orbits.

One condition we may state with certainty is that Trojan planet formation requires 
the presence of multiple planets within a system. 
Current observations find that $\sim 10$\% of 
observed planetary systems host multiple planets, a figure that is expected
to increase as detections of lower-mass bodies increase in number with the
recent and future
launches of newer, more sensitive space missions ({\it e.g.}~KEPLER, CoRoT).

Recent models of planetary formation have suggested different mechanisms by 
which Trojan planets may form. {\it In-situ} formation of a co-orbital 
companion, possibly through an enhanced local density of solids near the 
triangular Lagrange points of a massive planet has been suggested by
\citet{laughlin02}  and examined more recently by \citet{beauge07}. 
Other authors 
describe the capture of existing bodies into the $L_4/L_5$ points through
a variety of mechanisms: the collision of two objects near a triangular 
Lagrange point or capture into a Trojan orbit due to rapid mass accretion
by the primary planet \citep{chiang05}; rapid convergent migration between 
a single gaseous and multiple rocky planets, 
directly into a co-orbital resonance \citep{thommes05}; or the 
scattering of existing objects into the $L_4/L_5$ points either through
large scale, global planetary system instability, during which 
the Trojan asteroids in our Solar System may have  been captured 
\citep{morbidelli05}, or local gravitational scattering
during the violent relaxation of a closely packed planetary system
\citep{cresswell06, cresswell08}. 
If capture in a co-orbital resonance occurs during or prior to planetary
migration, the libration amplitude of the oscillations may be decreased; 
numerical simulations have shown this to occur for both gaseous discs
\citep{cresswell08} and planetesimal ones \citep{ford07}. This damping causes
initial horseshoe orbits to become tadpole orbits around one or other of the
$L_4$/$L_5$ Lagrange points. Three-dimensional hydrodynamic 
simulations indicate that co-orbitals
formed from planetary scattering typically lie in or near the disc midplane, 
and in any case the presence of a gaseous disc rapidly damps the 
inclination of a companion body into the same plane as its primary \citep{cresswell08}.

\begin{table*}[t]
\caption{\label{tab1} Table summarising the differences between each model run.}
\begin{center}
\begin{tabular}{lllllllll}
\hline\hline
Model & $m_{\rm{t, i}} $ & $m_{\rm{l, i}} $ & $f_{\rm{l}} $ & $\alpha$ & $M_{\rm{disc}}$ & $m_{\rm{t, f}}$ & $m_{\rm{l, f}}$ & $q_{\rm{p}}$ \\ 
 & ($M_{\oplus}$) & ($M_{\oplus}$) & & & & ($M_{\oplus}$) & ($M_{\oplus}$) & \\ \hline
1 & 15 & 15 & 5/3 & $5 \times 10^{-3}$ & 1 & 145 & 125$^{\mbox{\tiny{\it 1}}}$ & $0.862^{\mbox{\tiny{\it 1}}}$  \\ 
2 & 15 & 15 & 5/3 & $5 \times 10^{-3}$ & 0.5 & 141 & 172 & 0.817 \\ 
3 & 15 & 15 & 5/3 & $10^{-3}$ & 1 & $\sim 125$ & $\sim 100^{\mbox{\tiny{\it 2}}}$ & $\sim 0.75^{\mbox{\tiny{\it 2}}}$ \\ 
4 & 15 & 1 & 0 & $5 \times 10^{-3}$ & 1 & 185 & 1 & $5.40\times10^{-3}$ \\ 
5 & 15 & 10 & 5/3 & $5 \times 10^{-3}$ & 1 & 182 & 69 & 0.378 \\ 
6 & 10 & 15 & 5/3 & $5 \times 10^{-3}$ & 1 & 62 & 191 & 0.324 \\ 
7 & 20 & 10 & 5/6 & $5 \times 10^{-3}$ & 1 & 190 & 45 & 0.235 \\ \hline
\end{tabular}
\end{center}
\end{table*}

Models that produce co-orbitals  because of dissipation provided by 
a disc may also experience significant inward migration, 
which would increase the probability of the planets' detection.
Many studies investigating the possibility of detecting Trojan planets focus 
on transit observations \citep{rowe06, ford07, gaudi07} or a combination of 
transit and radial velocity methods \citep{ford06}, since 
additional physical information may be obtained in the event of 
discovering a body with Earth-like properties. Observational
constraints on Trojan formation will provide clues to the formation history 
of a planetary system, for example indicating that a period of violent
relaxation has occurred in a highly dissipative environment \citep{cresswell06, cresswell08},
or for short period systems differentiating between circularisation
near the star of an initially highly eccentric orbit versus migration through a gaseous disc.

In this paper we consider Trojan systems that have been formed within
a gaseous protoplanetary disc. Implicit in our assumption is that
the formation mechanism is the same as that proposed by 
\citet{cresswell06, cresswell08},
since the Trojan systems we consider are similar to those found as             
intermediary or end-states of the simulations in
\citet{cresswell08}.
We follow their evolution as they migrate and accrete gas from the disc, through
the epoch of gas disc dispersal, and finally in
the disc-free environment to determine the stability of such pairs throughout a global
history of the system.

This paper is organised as follows.
In Sect.~\ref{method}  we describe the models used, 
and in Sect.~\ref{results} we present our results.
In Sect.~\ref{conclusions} we conclude with a short discussion 
of the implications of this work for the detection of Trojan planets.

\section{Numerical Methods}
\label{method}

We use both hydrodynamic and modified $N$-body simulations in this paper.
We first discuss the hydrodynamic scheme in Sect.~\ref{hydro-method}, 
followed by the $N$-body method in Sect.~\ref{nbody-method}.

\subsection{Hydrodynamic method}
\label{hydro-method}

Due to the comparatively short inclination damping time, we have found that in
the presence of a protoplanetary disc most Trojan systems are quickly reduced 
to co-planarity (to within $<0.1^{\circ}$) and remain in that state in
the absence of any further influences \citep{cresswell08}. In order
to maximise the simulated time of our models on available hardware we therefore
elected to utilise two-dimensional simulations of the disc-planets system.

We use the NIRVANA hydrodynamic code in two dimensions using $(r,\theta)$ 
co-ordinates (see \citet{cresswell06} for details). The computational domain 
is given by $[2,12.5]$ AU in radius and $[0,2\pi]$ radians in azimuth. 
The number of  uniformly spaced grid cells 
used in the radial and azimuthal directions is
$N_r=300$ and $N_{\theta}=900$, respectively.
The disc is assumed to be locally isothermal with a constant aspect ratio
$h=H/r=0.05$, and the surface density profile is given by 
$\Sigma(r)=\Sigma_0 r^{-1/2}$.  For most models $\Sigma_0$ is normalised
such that the disc contains 0.04 $M_{\odot}$ of gas interior to 40 AU.
We adopt a standard $\alpha$-model for
the disc viscosity, such that the kinematic viscosity $\nu = \alpha c_s H$
where $c_s$ is the local isothermal sound speed \citep{shakura73}.
At the inner radial boundary we use a viscous outflow condition to
define the radial velocity \citep{pierens08}, such that
$v_r = \beta v_{\rm{visc}}$, where $v_{\rm{visc}}(r)=-3\nu / 2r$ is the inward
drift velocity of a steady-state accretion disc. $\beta$ is a free parameter
for which we choose $\beta=5$ in accordance with \citet{pierens08} and 
\citet{crida07}; however other nearby values of $\beta$ were tested and found 
to produce no significant differences in the resultant planetary systems
formed. The outer boundary is reflective
with a wave-damping function adopted between $r=10.5$ and $12.5$ AU to
minimize wave reflections \citep{cresswell06}.

We perform runs using a variety of initial planet masses and accretion rates (see
table \ref{tab1};
from left to right, columns represent: model number, initial mass of trailing
planet, initial mass of leading planet, accretion factor of leading planet,
viscous $\alpha$-parameter, disc mass (normalised against
model 1), final mass of trailing planet, final mass of leading planet, 
and final mass ratio of the two planets). 
The planets are initially in tadpole orbits in all models;
initial positions and velocities are taken from a simulation in 
\citet{cresswell08}, where a co-orbital pair has recently damped to stable,
co-planar tadpole librations. Although in each run the planets will be 
librating around slightly different points with different amplitudes, in each 
case the planets' orbits make the necessary adjustments within a few orbital periods,
aided by the damping action of the surrounding disc. The planets' orbits are 
evolved using a fifth-order Runge-Kutta method, and by calculating the 
torque from the disc on each planet; the torque from material located 
within one Hill radius of each planet is neglected from this torque 
calculation,  using a Heaviside step function.
A gravitational softening parameter
$\epsilon=0.5H$ is used when calculating the planet potentials. 
The time step is calculated by determining separate N-body and hydrodynamical
timesteps, and utilising the smaller value; further details are
provided in \citet{cresswell06}.

\begin{figure}[t]
\begin{center}
\includegraphics[width=7cm]{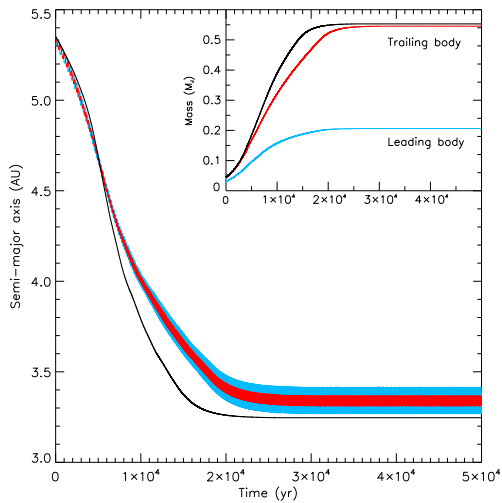}
\caption{\label{fig1} ({\it Main}) 
The semi-major axes of the leading (blue line) and
trailing (red line) planets in model 5; note the oscillations in $\Delta a$
are larger for the leading body, so that the red line appears superposed
above the wider blue line.
Stable, near-circular tadpole orbits are retained at all times.
The evolution of the same 15 M$_{\oplus}$ planet in isolation is shown for comparison (black line).
({\it Inset}) The planets' masses, in Jupiter masses, over the same time frame.}
\end{center}
\end{figure}

Accretion onto the planets is handled in a similar way to \citet{dangelo03},
with the density within 0.1 Hill radius of the planet reduced by a factor
$1-f\Delta t$ during each time step. As with 
\citet{paardekooper08}, we choose $f=5/3$ in most 
models, but other values are used for the accretion rate onto the leading
planet (labelled $f_{\rm{l}}$ in table~\ref{tab1}) in some cases. 

Each simulation progresses until one planet reaches a prescribed mass,
whereupon at each iteration the mass of the disc 
is reduced everywhere according to an exponential decay law with an
$e$-folding time of approximately 2550 years. 
This prescribed value is usually 0.5 Jupiter masses ($M_{\rm{J}}$), although
in some cases (models 1 \& 3) where both planets possess the same initial mass 
this value cannot be achieved prior to the planets migrating all the
way to the inner boundary. This occurs because systems with the same initial masses
usually show a reduced rate of accretion onto both planets as they
compete for gas.
In these cases a lower value is taken for the planet mass at which gas dispersal
is initiated. See 
table~\ref{tab1} for further details of each run.
After approximately a further 
$1.2\times 10^4$ years of disc mass reduction, 
disc forces are disabled completely (i.e.~the system is
purely $N$-body) to determine stability without the effect of the disc.
Although disc dispersal time scales are believed to be 2 -- 3 orders of
magnitude longer than those used here, the rapid decay we employ 
makes the transition from embedded to free planets computationally tractable.
Test simulations of resonant planets using HENC-3D (an $N$-body code adapted to
emulate the effects of hydrodynamic disc forces, see Sect.~\ref{nbody-method})
have shown that as long as the decay time scale is not near-instantaneous, the 
rate at which disc mass is removed is not responsible for determining whether 
the resulting system remains stable in the absence of a surrounding disc.
The time scale we use is longer than any of the natural dynamical time scales
involved in the problem, including the secular evolution time for the planet orbits,
suggesting that our gas dispersal algorithm should be reasonably successful in
capturing the process of slow disc dispersal.

\subsection{$N$-body method}
\label{nbody-method}

At the present time it is not possible to run hydrodynamic simulations
for sufficient time to simulate very large scale migration of
Trojan planet pairs. The simulations performed using NIRVANA included
a routine for modelling disc dispersal that normally switches on
when the planet semi-major axes have approximately halved from their
initial values. To simulate the
formation of short-period Trojans 
we use an $N$-body code adapted to emulate
hydrodynamic forces, HENC-3D (see \citet{cresswell08} for details).
The migration and eccentricity and inclination damping rates of HENC-3D
are founded on the analytic prescriptions of \citet{tanaka04} and
\citet{papaloizou00},
and are calibrated against NIRVANA to provide good approximations to the
migration and eccentricity (and inclination) damping rates; it utilises the
same fifth-order Runge-Kutta integrator as NIRVANA.
Since this $N$-body code uses semi-analytic terms to model the disc forces
rather than modelling the disc itself and calculating the resultant force,
mass accretion is simulated using a hyperbolic tangent function, tailored to each
model, such that planets reach the same final mass at the same time
as their counterparts in the hydrodynamic models. Independently of their
growth, after an appropriate time the disc forces are reduced to simulate
excavation of a gap and the reality that the type I forces modelled by HENC-3D
would be too large for a body that has accreted significant amounts of
gas. Both of these additions were tested against all the models
used by HENC-3D to ensure agreement with NIRVANA before further simulations
beyond NIRVANA's capacity for inward migration were conducted.
Disc dispersal was modelled to occur on the same time scale to that
described above for the NIRVANA simulations, and was mimicked by simply
reducing the disc forces using an exponential decay law.

\begin{figure}[t]
\begin{center}
\includegraphics[width=9cm]{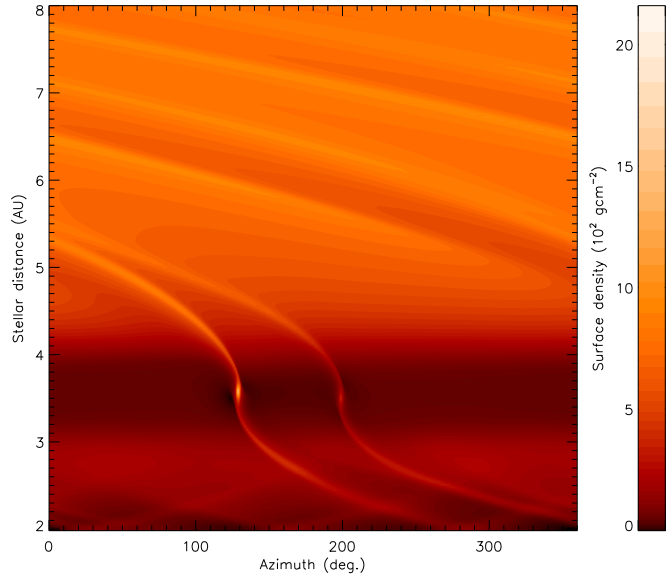}
\caption{\label{fig2} Surface density of the inner region of the disc in
model 5 at $t\approx16,700$ yrs.
The planets' masses at this time are
156 $M_{\oplus}$ (left) \& 65 $M_{\oplus}$ (right).}
\end{center}
\end{figure}

\section{Results}
\label{results}
\subsection{Orbital evolution} \label {evolution}
We begin by describing the orbital evolution of the Trojan systems.
Fig.~1 shows the evolution of the semi-major axes for the two planets
in model 5, whose initial masses were 15 and 10 M$_{\oplus}$, and final masses
after disc dispersal were 182 and 69 M$_{\oplus}$. The two planets migrate inward
at close to the standard type I migration rate as they accrete gas, and remain
stable throughout their evolution. 
The initial phase of inward type I migration occurs because of our adoption of
a locally isothermal equation of state. \citet{paardekooper08} have shown that
migration in a radiatively inefficient disc may be stalled or reversed, but
once gap formation ensues the corotation torques responsible for this 
should decrease, leading to long-term inward migration.
Also plotted in Fig.~1 is the semi-major
axis evolution of an isolated 15 M$_{\oplus}$ planet 
undergoing the same accretion process.
Comparison between
the migration rates shows that the Trojans begin to migrate slightly faster
early on in their evolution as their masses grow and type I migration rates
increase, but as gap formation starts to occur, and disc dispersal is switched
on, their migration begins to slow before finally halting after the gas disc
has been completely removed. 
The inset to Fig.~\ref{fig1} shows that the growth of one planet is
accompanied by the inhibition of the accretion rate of its companion; 
however the larger body does not accrete at the same rate as the equivalent
isolated planet, with its Trojan partner removing some of the material in the
horseshoe region.
Divergence of the masses of the Trojan
planets during migration means that their individual migration rates should also
diverge, but the 1:1 resonance is maintained during migration,
causing the planets to migrate inward in lock step. The long term evolution
in the absence of gas consists of the planets orbiting stably at
their mutual $L_4$/$L_5$ Lagrange points, with small amplitude librations
in semi-major axes caused by the libration associated with the tadpole orbits.
Although initially damped to small librations by the disc, 
as the planets migrate inwards the amplitudes of these librations increase 
slightly, but always remaining firmly within the tadpole regime over the radial
distances covered here  (but see Sect.~\ref{short} for a discussion 
about the evolution during large-scale migration).
The orbital eccentricities remain small during migration, increasing slightly
upon gas disc removal but always remaining below $e = 0.02$.

 Figure \ref{fig2} shows the surface density of the inner disc in model 5 after
approximately $1.67\times10^4$ years, shortly before disc dispersion is
applied. By this time the two bodies have cleared a significant gap in the 
disc, though the larger body dominated the local dynamics. Some wave reflection
is visible at the inner boundary; 
the planets in different models ceased migration
(due to disc dispersal) at varying distances from the inner boundary, some
close to the depleted inner edge of the disc and within range of the reflected 
waves, and some not, but no discernible effects in the evolution of any Trojan
pair was observed as a result. Indeed, with the exception of the final
orbiting radius, the orbital evolution and surface density profile of each
model was very similar.

We note here that our adoption of a `viscous outflow' boundary 
condition prevents rapid loss of the inner disc as is observed when open 
boundary conditions are used. Our adoption of an outflow velocity maximum
that is five times larger than the nominal viscous rate, however, can still
cause the inner disc to deplete a little too rapidly. Potentially this could
affect the outcome of our results, and so we have run tests in which the
outflow velocity maximum was varied with both larger and smaller values used.
We find that our results are insensitive to the adopted values.

With only minor variations in description,
this summary applies equally to all of models 1 -- 7, indicating that
Trojan planets are generally stable and insensitive to their relative masses
and local disc conditions, at least within the parameter ranges considered
 in these hydrodynamic simulations.
In agreement with \citet{cresswell06,cresswell08} Trojans also remain stable
throughout significant migration, including during the rapid migration which
occurs as the planets grow but prior to gap formation.
The onset of gap formation during the evolution of model 5 is shown in Fig.~2,
which displays a snapshot of the disc surface density profile.
The position of the planets is clearly visible within the gap
caused by the combined effects of gas accretion and the tidal torques
induced by the trailing planet, whose mass at this point in time has  
almost reached 0.5 M$_{\rm J}$.

\subsection{Planetary mass growth}
\subsubsection{Equal initial planet masses}
In models 1 -- 3 the planet masses at the beginning of the simulations
were all 15 M$_{\oplus}$, with leading and trailing planets having the 
same mass. In model 1 the migration rate of the planets was such that
they both began to approach the inner boundary of the computational
domain prior to either of them reaching 0.5 M$_{\rm J}$, the mass at
which gas disc dispersal is switched on. Therefore, in this model,
gas disc dispersal was initiated when the most massive planet had reached
$m_t=0.41$ M$_{\rm J}$. The evolution of the planet masses is shown in
Fig.~3, and the final masses are listed in table.~1. For the case of
model 1 it can be seen that the planet masses remain similar during the
evolution, but there is a slight divergence with the trailing planet
being the larger one.

Model 2 used a very similar set up to model 1, except the disc surface
density everywhere was halved. This caused the migration to slow, and
allowed the planetary growth to proceed until one of the planets
reached 0.5 M$_{\rm J}$, after which gas disc dispersal was initiated.
Interestingly, we find that the planet masses in this case also
remain similar during the evolution, but in this instance
the leading planet becomes the larger one, in contrast to the
situation in model 1. This indicates that in general, when Trojan planets
of equal mass begin to accrete gas at the same time their masses will
remain similar, but some level of divergence will occur due to stochastic
changes in the accretion rates as the system evolves. Our simulation sample
is too small to indicate any underlying bias between the leading or trailing planets
becoming the more massive, but does show that either the leading or trailing
planet may grow faster. Once the planetary mass ratio diverges
from unity, then the simulations indicate that the divergence tends to
accelerate, for reasons discussed in the following subsection.

Model 3 used very similar parameters to model 1, except the disc 
viscosity was reduced from $\alpha=5 \times 10^{-3}$ to $10^{-3}$.
During the process of gap opening this allows the gap to deepen
more quickly, since material is less able to viscously diffuse
into the gap. 
 With the added effect that the planets are competing for the same material
to fuel their growth, 
the result is that the planets grow  significantly slower and
end up with smaller masses, as shown in table.~\ref{tab1}.
The slow growth of these planets caused them to migrate close to
the inner boundary before reaching 0.5 M$_{\rm J}$, and the simulation
was simply halted at this point.

\begin{figure}
\includegraphics[width=8.5cm]{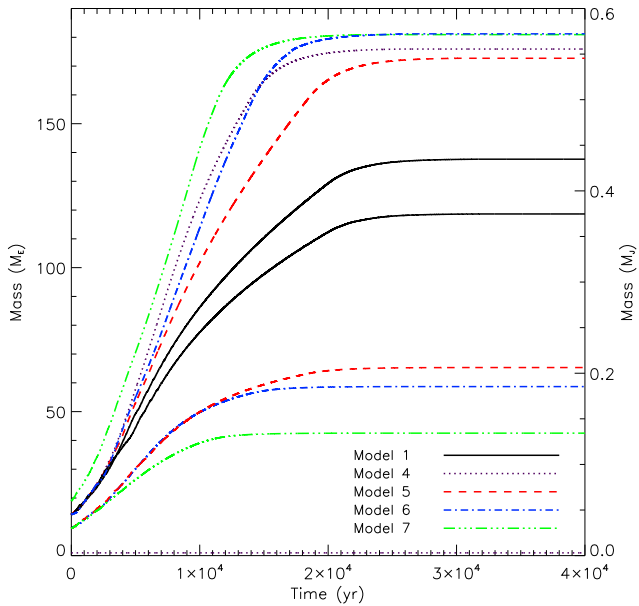}
\begin{center}
\caption{\label{fig3} Planetary masses for a selection of models; all planets
shown had reached their final mass before $4\times10^4$ yrs.}
\end{center}
\end{figure}

\subsubsection{Unequal initial planet masses}
Models 4 -- 7 all started with unequal initial planet masses.
In model 4, the trailing planet had a mass of 15 M$_{\oplus}$
and was able to accrete at the normal rate, but the leading planet
had a mass of 1 M$_{\oplus}$ and was unable to accrete since this
mass is below the critical core mass for gas accretion \citep{pollack96}.
This model simply serves to show that a Trojan system which forms
with an Earth-like planet component can evolve stably to the stage
where the primary component is a gas giant. 

Models 6 and 7 have initial planet masses of 10 and 15 M$_{\oplus}$,
and differ only in which of these planets is leading or trailing.
Prior to gap opening, the larger Hill radius of the more massive
planet causes this larger planet to have a higher accretion rate,
causing the initial mass difference to grow.
As gap formation begins to occur,
viscous gas flow into the gap from the main body of the disc allows the
more massive planet with the largest Hill radius to dominate the
accretion process, since it is able to intercept this gas before the
smaller planet can. 
The dominating body proceeds to accrete disc mass while depriving its less
massive companion of material, and the mass ratio between the two planets
moves successively further from unity; this is demonstrated in the inset
to Fig.~\ref{fig1} which shows the evolution of planet
masses for model 5, 
and in Fig.~\ref{fig3} which shows the evolution of both models. 
With the formation of a deep gap in the disc, the
accretion rate of the smaller planet falls to near-zero while that of the
larger body follows linear growth (until disc mass reduction begins).
In our simulations, the mass of the 
smaller body is typically capped at 50 -- 70 $M_{\oplus}$ when the
initial planet masses are unequal, while the 
accretion rate of the larger body is largely unaffected by the presence of a 
companion. We performed additional test calculations with initial conditions
similar to models 5 and 6, but with gas disc dispersal switched off.
In those models we also found that the accretion of gas by the 
less massive planet was also diminished when gap formation
by the heavier planet ensued, resulting in final masses of between 50 -- 70
M$_{\oplus}$ for the lighter planet. This shows that these final masses
are not a product of the gas disc dispersal, but because the lighter
planet becomes starved of gas when gap formation occurs.

In model 7 the trailing planet is twice as massive initially
as the leading planet (20 versus 10 M$_{\oplus}$).
In addition the lower mass planet had a slower gas accretion rate,
as might be expected from detailed models of gas accretion onto
solid planetary cores \citep{pollack96}. 
 In light of the naturally diverging accretion rates among all the Trojan pairs
we studied, the differing accretion prescriptions merely exacerbates the
effect. Consequently 
the results of this simulation are
rather unsurprising: the initially unequal planet mass ratio becomes
more unequal during the evolution, and the system remains stable 
for the duration of the simulation.

It is worth commenting that our models of accreting Trojans
indicate that two accreting
ice and rocky planetary cores are 
able to clear a gap
in the disc quicker than an isolated planet.
For example, after $5\times10^3$ years the azimuthally-averaged surface density
at the radial location of the planet ($\Sigma_{\rm{p}}$) is observed to be
15\% lower in model 5 than in the case of an isolated $15 M_{\oplus}$ planet,
and 27\% lower in model 1. The gap in both cases is also excavated quicker,
having up to a 50 -- 60\% lower value of $\Sigma_{\rm{p}}$ after
0.5 -- $3 \times 10^3$ years. One interesting by-product of this
is that the Trojan pairs will spend less time undergoing type I migration
than an isolated, accreting planetary core, helping prolong their life time
in the disc.

\subsection{Long term stability}
We have run all models 1 -- 7 for $10^9$ years in the absence of gas,
and find all systems to be stable, with no long term increases in eccentricity
or libration amplitudes in semi-major axes. These results indicate
that Trojan planet systems consisting of at least
one gas giant planet and possibly two, 
at the distances from the star considered previously, 
are extremely stable during and after their formation.

\subsection{Formation of short-period Trojans}
\label{short}
 The Trojans in our NIRVANA simulations are limited to migrating 
to within a few AU of the star, having formed with semi-major
axes $\simeq 5.3$ AU.
To observe the stability of Trojan planets as they migrate
to much smaller radii, we use
the HENC-3D $N$-body code described in Sect.~\ref{nbody-method}, 
which provides a reasonable approximation of the
hydrodynamic forces at work.

The models are run as follows: for each of the models 1, 4 \& 5 described
in table 1, the relevant Trojan 
pair is allowed to migrate all the way towards the star without
disc dispersal. Each run is then
repeated with disc dispersal simulated once the the semi-major axis of the
more massive planet reaches 0.5, 1.0 or 1.5 AU. These runs are 
repeated a further
three times for all cases using a small random perturbation of the initial 
conditions, including a vertical displacement not possible in the 2D NIRVANA
runs, to generate a set of four distinct tadpole pairs for each initial
mass ratio and stopping distance.

As noted in the NIRVANA models (Sect.~\ref{evolution}, see Fig.1 in particular),
after a brief initial period of decrease,
the amplitude of the tadpole librations
slowly increases as the planets migrate inward. 
Those planets stopped by simulated disc dispersal at distances of
$\approx 0.5$ AU from the star always survived in tadpole orbits, 
with minimum/maximum ($\theta_{\rm{min}}$/$\theta_{\rm{max}}$) separations
between the two planets of up to $\theta_{\rm{min}} \gs 25^{\circ}$ and
$\theta_{\rm{max}} \ls 130^{\circ}$.
However, the libration amplitude of those planets allowed to migrate further 
continues to increase until the system becomes unstable. 
Dissociation of the co-orbital pair is usually first signified by a shift from
tadpole to horseshoe orbits, which 
typically last for less than $3\times10^4$ yrs (although 
horseshoe orbits lasting over $10^5$ were also observed). This is finally
followed by either collision (76\% of cases) or scattering onto distinct 
orbits. In one instance differential migration caused the two bodies
to form a 5:3 mean motion resonance (MMR) after scattering.

This amplitude increase and eventual destruction of the co-orbital structure
was observed regardless of the initial mass ratio and, in further tests, the
migration and eccentricity damping rates of the disc. After starting their
migration from 5.3 AU in our simulations, in all cases the 
instability took effect between 0.5 -- 0.1 AU.

We also performed numerical experiments which involved:  
{\it (i)}.\hspace{1mm} placing planets on initial tadpole orbits with smaller 
libration amplitudes; {\it (ii)}.\hspace{1mm}
retaining the original libration amplitudes but placing planets
closer to the star initially. Both of these changes allow 
planets to remain on tadpole
orbits while migrating very close to the star (i.e. Trojans with
semi-major axes $\le 0.1$ AU), raising the possibility 
that `hot Trojans' containing a gas giant planet component
may form and survive.  
The potential observability of such systems is therefore highly
dependent on the initial orbits of the planets when the Trojan systems
are established, and the ability of the disc to
damp large initial librations directly after the Trojan systems have
formed.

We note that, despite the abundance of Trojans ($\sim 30$\%) in the simulations
of \citet{cresswell08}, the disruption of these systems
during large-scale migration was not always detected there.
After re-examining that data we note that this stability
was found to occur only in Trojan systems which were also in a mean motion
resonance with another body. Apparently this additional body is able to provide
a stabilising effect, allowing a Trojan system to migrate closer to the
central star. We note however that an additional body did not always provide
a stabilising influence, and similar scattering behaviour was observed in some
of the runs presented in \citet{cresswell08}.

\section{Discussion and conclusions}
\label{conclusions}

We have investigated the evolution of so-called tadpole or Trojan planets,
whose masses are initially in the Neptune-mass (or Earth-mass)
range, in the presence of
a gas disc from which they are able to accrete to become gas giant planets.
For such an isolated pair,
previously damped onto coplanar, near-circular orbits by the disc after
the co-orbital configuration was formed
(as found by \citet{cresswell08}), 
we find that each pair remains stable during rapid type I migration and
gas accretion from the disc. This period of stability also extends
through the epoch of gap formation in the disc, and  
the period of gas disc dispersal, resulting eventually in Trojan systems 
which are stable for $10^9$ years.

For Trojan systems of planets with masses which were initially equal, we
found that gas accretion resulted in final gas giant planets with similar
but not equal masses. For systems with unequal initial masses, we find
that the initially more massive body dominates the accretion and approaches
the Jovian mass
as if in isolation. The lower mass planet is effectively starved of gas
when gap formation ensues, causing its final mass to not exceed
70 M$_{\oplus}$, at least for the parameters we have considered. 
This suggests that Trojan planetary systems in nature consisting
of one component which is a gas giant are likely to have a second
component which is considerably smaller in mass.

Our results indicate that if a mechanism exists to form
significant numbers of Trojan planets, such as the violent
relaxation of a planetary system in the presence of a
substantial gas disc \citep{cresswell06, cresswell08, morbidelli08},
then Trojans should be observed in nature as they are
expected to be very stable,  provided they have not undergone
very large-scale migration.
The simulations indicate that large disparities in mass are expected
between the Trojan pairs, 
but detection of co-orbital
systems with planetary masses in the Jupiter- and Saturn-mass ranges, 
or similarly-scaled Trojan pairs, 
will indicate a prolonged period of simultaneous co-orbital growth.

\begin{figure}[t]
\begin{center}
\includegraphics[width=9cm]{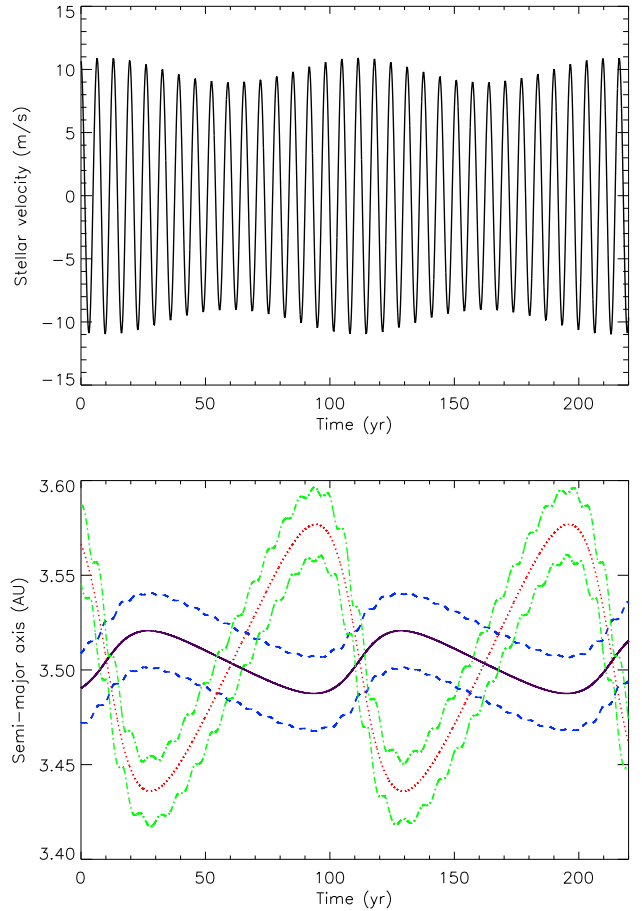}
\caption{\label{fig4} ({\it Top}) The radial velocity curve of a $1 M_{\odot}$
star bearing a Trojan planetary pair, taken from the final state of model 7.
({\it Bottom}) The semi-major axes of the orbiting planets, with masses
$190 M_{\oplus}$ (continuous indigo line) and $45 M_{\oplus}$ (dotted red
line), over the same time. The instantaneous pericentres and apocentres of each
planet are also shown.}
\end{center}
\end{figure}

 We also find that inward migration produces an increase in the libration
amplitude of Trojan planets about their mutual Lagrange points, 
eventually leading to instability and 
collision or planet-planet scattering if large-scale migration has taken
place. This may have unfortunate implications
for the detection of short-period Trojan planets containing
one component which is a gas giant, since Trojan systems formed at $\simeq 5$ AU
and with initial librations extending $\gs 15^{\circ}$ 
in azimuth from the $L_4$/$L_5$ point
are expected to become unstable after they have migrated
to within a few tenths of an AU of the central star. 
We expect that if such Trojan planets are observed, they will have 
entered into a co-orbital relationship somewhat closer to the star than in
the simulations considered here,
or have been damped by the disc onto --- or otherwise formed with --- 
much smaller tadpole librations before the opening of a gap. 
The simulations of \citet{cresswell08} produced Trojans that cover both of these
possibilities, with initially large librations damped by the disc 
to a few degrees within a few $\times 10^4$ years. The formation of a 
planet at the Lagrange point (e.g. \citet{beauge07}) may also produce
sufficiently small librations to survive significant migration.
Trojan pairs in mean motion resonance with a third body may also be subject
to sufficiently small librations to survive large-scale migration.

As a means to reduce the effects of type I migration,
\citet{thommes05} suggested that gap-opening planets may provide a moving 
barrier by capturing fast migrators into mean motion resonances following
behind the giant planet; in some instances multiple terrestrial bodies were
observed to share the same MMR. Co-orbital capture of a terrestrial body with
an (eventual) gas giant itself, in some cases even before gap-opening has 
begun, may serve the same function and act as a natural (and stable, in a 
post-disc environment) extension of this 
`stacking' behaviour to include the 1:1 resonance with the giant planet itself.
However, this does not address the prior issue of achieving a 
gap-opening planetary mass.

A sample radial velocity (RV) curve produced by a star orbited by a Trojan pair 
is shown in the top panel of Fig.~\ref{fig4}. The tadpole motions of Trojan 
planets produce a regular variation in their semi-major axes, which is clearly
visible in the star's RV profile, amounting in this instance to a regular
modulation in the curve's amplitude of around 20\%. In the given model the 
associated `beat' has a period of approximately 100 yrs; 
if the planets were oscillating around 1 AU, this period would be only 17 yrs, 
with the peak amplitude of stellar radial velocities being modulated
between 17 and 20 ms$^{-1}$; 
at 0.1 AU, this becomes $\sim$180 days and variations in the peak
stellar radial velocity occurring between 53 and 64 ms$^{-1}$.
Unfortunately the amplitude of this variation depends on the
history of the system, since tadpole planets with minimal
libration will show a very small variation, whereas systems with
larger amplitude libration will show larger modulation of the
radial velocity, so it is not a straightforward matter to define the
expected amplitude of variation as a simple function of system
parameters. The general shape of the RV curve, however, is
a tell-tale sign of a Trojan system, and arises simply because
the centre of mass of the system around which the star orbits
varies with time as the planets librate around the $L_4$/$L_5$ points.
When the planets are at their furthest distance apart in
azimuth the system's centre of mass lies closest to the central star,
and its radial velocity amplitude is minimised. Conversely the centre of mass
is at a maximum distance from the star when the planets are at closest
approach, causing the radial velocity to be at a maximum. Our simulations
suggest that variations in the radial velocity
amplitude on the order of 10 -- 20 \% are expected in Trojan systems
over time scales of about 20 orbital periods.
This should allow short period gas giants with hot-Neptune Trojan
companions to be discovered from radial velocity surveys, and
if they exist, hot Neptunes with Trojan super-Earth companions. 
As mutual inclinations
are expected to be small due to gas disc inclination damping, it
is also possible that current and forthcoming transit surveys such as
CoRoT and KEPLER will discover Trojan planet systems.

\acknowledgements
The simulations presented in this paper were performed using 
the QMUL HPC Facility purchased under the SRIF initiative.
We thank Gareth Williams for several interesting discussions 
on the subject of co-orbital dynamics, referee Aur\'{e}lian Crida
for the constructive criticisms which helped improve this paper,
and Tristan Guillot for additional comments which further enhanced 
our study.

\bibliographystyle{aa}

\bibliography{trobib}

\end{document}